\documentclass[twocolumn,prl,a4paper,showpacs,floatfix]{revtex4}
\usepackage{graphicx}

\newcommand{\ud}{\mathrm{d}}
\newcommand{\be}{\begin{equation}}
\newcommand{\ee}{\end{equation}}
\newcommand{\Ef}{E_{F}}
\newcommand{\Eperp}{E_{\bot}}

\newcommand{\Ez}{E_{z}}
\newcommand{\wz}{\omega_{z}}
\newcommand{\wperp}{\omega_{\perp}}
\newcommand{\w}{\omega}

\newcommand{\Pz}{P_{z}}
\newcommand{\px}{p_{x}}
\newcommand{\py}{p_{y}}
\newcommand{\pz}{p_{z}}

\newcommand{\zt}{\tilde{z}}

\newcommand{\pt}{\tilde{p}_{z}}

\begin{document}

\draft

\title{Insulating Behavior of a Trapped Ideal Fermi Gas}

\author{L.Pezz\`e$^{1}$, L.Pitaevskii$^{1,3}$, A. Smerzi$^{1,4}$,
S.Stringari$^{1}$, G. Modugno$^2$, E. DeMirandes$^2$, 
F. Ferlaino$^2$, H. Ott$^2$, G. Roati$^2$, M. Inguscio$^2$}
\affiliation{ 
$^1$ Istituto
Nazionale per Fisica per la Materia BEC-CRS and
Dipartimento di Fisica, Universit\'a di Trento, I-38050 Povo, Italy\\
$^2$ LENS and Dipartimento di Fisica, Universit\'a di Firenze, and INFM
Via Nello Carrara 1, 50019 Sesto Fiorentino, Italy\\
$^3$ Kapitza Institute for Physical Problems, 117334 Moscow, Russia\\
$^4$ Theoretical Division, Los Alamos National Laboratory, Los Alamos, NM 87545,
USA}
\date{\today}

\begin{abstract}
We investigate theoretically and experimentally the center-of-mass
motion of an ideal Fermi gas in a combined periodic and harmonic
potential. We find a crossover from a conducting to an insulating
regime as the Fermi energy moves from the first Bloch
band into the bandgap of the lattice. The conducting regime is
characterized by an oscillation of the cloud about the potential
minimum, while in the insulating case the center of mass remains
on one side of the potential.
\end{abstract}

\pacs{PACS: 03.75.Ss Fermi gas, 03.75.Lm BEC in optical lattices,
63.20.Pw localized states, 05.45.-a Non linear dynamics}
\maketitle

Recent experiments have demonstrated the possibility of
reaching the superfluid regime with trapped fermionic atoms
\cite{jin,bec}. In this context,
the combination of trapped cold fermions and optical lattices
\cite{Modugno2003}, is very promising. On the one hand it
represents the natural analogy with superconductivity of electrons
in crystal lattices, while on the other it might be a
additional tool to manipulate the
interaction properties of the system \cite{hofstetter}, and to
investigate the superfluid phase.
Atomic Fermi gases in lattices are
interesting also in view of observing novel quantum phases that
have been predicted for interacting Fermi \cite{fermi} and
Fermi-Bose systems \cite{fermibose}.

In this work we investigate  a Fermi gas of non interacting atoms
trapped in a harmonic field and subjected to a one-dimensional
optical lattice. We study both theoretically and experimentally the
center-of-mass dynamics of the system after a sudden displacement
of its equilibrium position along the lattice. We find that
the statistical distribution makes the dynamical
properties of such system highly non trivial. The most dramatic
new feature is that the system can exhibit a conducting or an
insulating behavior, depending on whether the Fermi energy lies in
the first Bloch band of the lattice or in the gap region. In the
first case the cloud oscillates symmetrically in the harmonic
potential, while in the second case it remains trapped on one side
of the potential. By tuning the width of the lattice band we
observe the full crossover between the two different regimes.

{\em Semiclassical Model.} Let us first introduce our
theoretical model to describe the system. Neglecting the
interatomic collisions, the many-body
Hamiltonian can be simply written as a sum
of single particle Hamiltonians:
\begin{eqnarray} \label{spH}
H_0 &=& \bigg( \frac{\pz^2}{2m}+\frac{1}{2}m \wz^2 z^2+s \, E_R
\sin^2(2\pi/\lambda z)\bigg)
+\nonumber\\
& & \bigg( \frac{\px^2+\py^2}{2m}+\frac{1}{2}m \wperp^2 (x^2+y^2)\bigg) ,
\end{eqnarray}
where $s$ is a tunable dimensionsless parameter, $E_R={\hbar^2
\over {2 m} }\big(\frac{2 \pi}{\lambda}\big)^2$ is the recoil
energy and $\lambda$ is the wavelength of the laser creating the
lattice. Since the harmonic oscillator length is typically
much larger than the lattice spacing $d=\lambda/2$ we can
study the system in the semiclassical approximation.

We first concentrate on the $z$-component of the Hamiltonian
(\ref{spH}). In the limit of a vanishing harmonic field
($\omega_{z} \to 0$), the eigenfunctions of the stationary
Schr\"odinger equation are Bloch waves. The eigenenergies
$\varepsilon_n(\pz)$ have the typical Bloch band structure as a
function of the quasimomentum $\pz$, defined in the first
Brillouin zone ($-\pi\hbar/d \leq \pz \leq +\pi\hbar/d$), and of
band index $n$ (in the following we consider only the lowest
energy Bloch band $n=1$). In a semiclassical approach, we can
incorporate the effects of the periodic potential modifying the
kinetic part of the Hamiltonian: $\frac{\pz^2}{2m}\to
\varepsilon(\pz)$, while the harmonic confinement generates a
driving field. Therefore, Eq. (\ref{spH}) can be
approximated as: \be \label{spH2} H_0=
\varepsilon(\pz)+\frac{1}{2}m\wz^2z^2 +
\frac{\px^2+\py^2}{2m}+\frac{1}{2}m \wperp^2 (x^2+y^2). \ee

The effects of a sudden small displacement $z_0$ of the center of
the harmonic trap can be included adding the perturbative term
$H_{pert}(z,t)=m \wz^2 \Theta(t) z_0 z$, where $\Theta(t)$ is the
unit step function. A first insight in the behavior of the system
is provided by the study of the linear dynamics in a 1-D
configuration.

\begin{figure}
\includegraphics[width=9 cm]{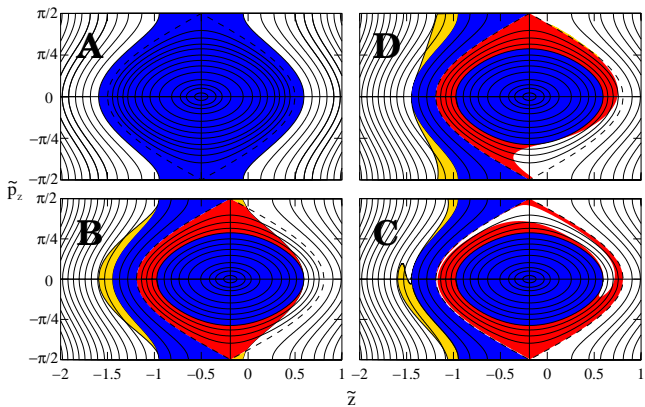}
\caption{\small{Phase trajectories for a trapped 1-D Fermi gas in
a lattice at $T=0$, just before and after the displacement of the
trap (figs. A and B, respectively), and their dynamical evolution
(figs. C and D). The ordinate and abscissa are in units of $\pt
\equiv \pz d/2 \hbar$ and $\zt \equiv\sqrt{m \wz^2 z^2/4 \delta}$.}} \label{phase}
\end{figure}

{\em 1-D Analysis.} The physics becomes particularly clear in the
semiclassical phase space, as shown in Fig.~\ref{phase}. The lines
are isoenergetic single particle orbits; they belong to two
different classes, separated by the dashed orbit, which has an
energy 2$\delta$, corresponding to the width of the first Bloch
band. The orbits are close when their energy is in the band, i.e.
the particles oscillate around the trap minimum. The orbits are
instead open when their energy is in the gap, because the
quasimomentum is reversed at the band edges. Particles in these
orbits oscillate on the sides of the harmonic potential,
performing the analogous of Bloch oscillations in a linear
potential. In general the oscillation frequency of both
kinds of orbits shows a strong dispersion with energy.

As shown in Fig.~\ref{phase}A, the Fermi gas at $T=0$ uniformly
fills the phase space region with energy below $E_F$. A sudden
displacement of the center of the harmonic potential corresponds
to a shift of the center of the phase space (see
Fig.~\ref{phase}B). The blue region contains particles that are
still in equilibrium in the new configuration of the trapping
field. The red and yellow regions, containing particles moving on
close and open orbits respectively, are instead out of equilibrium
and give rise to a collective dipole motion. The phase space
region opens and melts during the dynamics as a consequence of the
energy dependence of the single particle oscillation frequency
(see figs. C, D), yet leaving constant the phase space volume
(because of the Liouville theorem) and therefore preserving the
Pauli principle. To one hand, the red orbits dephase on a longer
time scale with respect to the yellow ones. Therefore,
the relaxation and the frequency of the oscillation mode are
dominated by the particles moving around the center of the phase
space, in the red region. To the other hand, the yellow orbits
remain open, trapping the center of mass of the system on one side
of the harmonic potential. Notice that in this 1-D configuration,
the damping of the oscillation disappears in the linear limit
(small initial displacement) at $T$=0.

{\em 3-D Analysis.}
The semiclassical phase-space distribution $f(\vec r, \vec p,t)$
is governed by the Liouville equation
\begin{equation}
\label{LP}
\frac{\partial f}{\partial t}+ \frac{\partial H}{\partial \vec{p}}
\frac{\partial f}{\partial \vec{r}}-\frac{\partial H}{\partial \vec{r}}
\frac{\partial f}{\partial \vec{p}} =0
\end{equation}
where $H=H_0+H_{pert}$. Such an equation correctly describes the
quantum dynamics up to orders of $O(\hbar^2)$ \cite{Wigner1}. For
small displacements, we can linearize the distribution function:
$f(\vec r, \vec p, t)=f_0(\vec r, \vec p)+g(\vec r, \vec p, t)$,
where $f_{0}(\vec r, \vec p)=\Big(\exp\big(\beta(H_0(\vec r, \vec
p)-\mu(T))\big)+1\Big)^{-1}$ is the fermionic solution of the
stationary Liouville equation for temperature $T$
relative to $H_0$, and $g(\vec r, \vec p, t)$ is a small
time-dependent correction induced by the perturbation
$H_{pert}(z,t)$. The dipole
oscillations evolve according to
: \be \label{Zt3d}
\frac{\langle \, \widehat{z}(t) \, \rangle}{z_0}=-1+\frac{2 m
\wz^2}{\pi} \int_{0}^{+\infty} \ud \w\, \bigg(\frac{1-\cos(\w
t)}{\w}\bigg) \Im\big(\chi(\w)\big). \ee where
$\Im\big(\chi(\w)\big)$ is the imaginary part of the dipole
response function $\chi(\w)$. In order to calculate it we solve
Eq.(\ref{LP}) in the linear regime by the general method
illustrated in \cite{Brink}, by using the energy of motion in
$z$-direction $\Ez=\varepsilon(\pz)+\frac{1}{2}m \wz^2 z^2$ as an
independent variable instead of $\pz$. The imaginary part of the
linear response function then becomes {\setlength\arraycolsep{2pt}
\begin{eqnarray} \label{imm3d}
\Im \big( \chi(\w) \big)&=&-\frac{1}{(\hbar \w_{\bot})^2}\frac{\pi}{\hbar} \sum_{n=1}^{+\infty}
\int_{0}^{+\infty} \ud \Ez \,\delta \big(\w-n \w_{0}(\Ez)\big) \, n {}\nonumber\\
& & {}\bigg[ \int_{0}^{+\infty} \ud \Eperp \, \Eperp
\frac{\partial f_{0}(\Ez+\Eperp, T)}{\partial E_{z}}
 \bigg] Q^2(n,\Ez) \nonumber\\
\end{eqnarray}}
where \be \label{Q} Q(n,\Ez)=-\frac{1}{n
\pi}\int_{z_{1}(\Ez)}^{z_{2}(\Ez)}\ud z \, \sin\Big(n\pi
\frac{\omega_{0}(\Ez)}{\omega_{0}(\Ez,z)}\Big) \ee and
\be
\label{w0E} \omega_{0}(\Ez,z)=\pi \Bigg(\int_{z_{1}(\Ez)}^{z}
\frac{\ud \zeta} {\big( \partial \varepsilon(\Pz)/\partial\Pz
\big)_{\Pz=\Pz(\zeta,\Ez)}}\Bigg)^{-1}. \ee
In the above equations
$\omega_{0}(\Ez)=\omega_{0}(\Ez,z=z_2(\Ez))$ is the single
particle frequency, $z_{1}(\Ez)$ and $z_{2}(\Ez)$ are the
inversion points of the motion (i.e. the maximum and the minimum
positions reached by the particle of energy $\Ez$). Notice that,
due to the 3-D nature of the system, all values $E_z \leq E_F$
contribute to the response function (\ref{imm3d}) even at zero
temperature, where the integral inside the square brackets becomes
$(\Ez-E_F)\Theta(\Ef-\Ez)$. As a consequence a damping due to
dephasing is in general present also in the linear limit, which is
different from the 1-D case. As shown by the Eqs.
(\ref{Zt3d}-\ref{w0E}), in order to calculate the dipole motion we
need to know explicitly the Bloch energy $\varepsilon(\pz)$ as a
function of the quasimomentum. In tight binding approximation
(large lattice height, corresponding to small tunnelling between
the potential barriers) the first band is $\varepsilon(\pz)=2
\delta \sin^2(\pz \lambda/4 \hbar)$ where $\delta\equiv \delta(s)$
\cite{Meret} is a decreasing function of the lattice
height. With this choice for the dispersion the frequency
$\w_0(\Ez)$ can be expressed in terms of elliptic integrals
\cite{Abramowitz} and we can calculate analytically the quantities
of interest. In regimes out of tight binding we use a
parametrization of the energy band \cite{banda}.

{\em Experiment.} We realize experimentally the system by using a
Fermi gas of $\mathrm {^{40} K}$ atoms. The sample is prepared in
the $F=9/2$, $m_F=9/2$ state and sympathetically cooled
\cite{Riboli} to quantum degeneracy in a magnetic trap with
frequencies $\w_z=2\pi \times 24$ s${^{-1}}$ and $\w_{\perp}=2\pi
\times 275$ s${^{-1}}$. Subsequently a one-dimensional optical
lattice is adiabatically superimposed along the weak axis of the
trap ($z$ direction) \cite{Modugno2003}. The lattice is created by
a laser beam in standing-wave configuration, with wavelength
$\lambda=863$ nm. The lattice height can be adjusted in the range
$U=0.1-8 E_R$, where $E_R/k_B=317$ nK. The typical Fermi gas is
composed by $25\,000$ atoms at a temperature that can be varied
between $0.2 \, T_F$ and $T_F$, where the Fermi temperature is
$T_F \approx 300$ nK.
\begin{figure}
\includegraphics[scale=0.57]{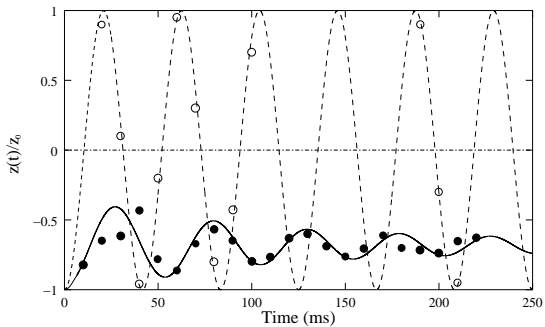}
\caption{\small{Dipole oscillations of the Fermi gas of $^{40}$K
atoms at $T=0.3\,T_F$ in presence (filled circles and full line)
and in absence (empty circles and dotted line) of a lattice with
height $s=3$. The lines are the theoretical predictions, the
circles are the experimental results. The horizontal dot-dashed
line represents the trap minimum.}}\label{oscliberaexp}
\end{figure}
The system is brought out of equilibrium by displacing the
magnetic trap minimum along the lattice. The typical displacement
is $z_0$=15 $\mu$m, much smaller than the 1/$e^2$ radius
of the cloud (110 $\mu$m) \cite{nota1}. After a variable evolution
time in the trap the atoms are released from the combined
potential. We detect the position of the center of mass of the
cloud by absorption imaging after a ballistic expansion of 8 ms.

\begin{figure}
\includegraphics[scale=0.7]{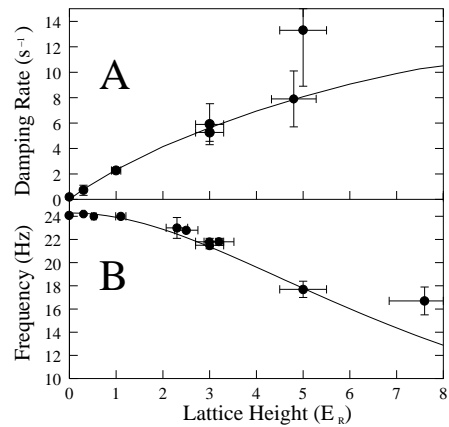}
\caption{\small{A) Comparison between theory (line) and
experiments (circles) for the damping rate of the dipole
oscillations of the Fermi gas as a function of the lattice height.
B) Oscillation frequency of the Fermi gas as a function of the
lattice height. The line is the expectation for a particle
oscillating at the band bottom. }}\label{dampingexp_articolo}
\end{figure}

{\em Results.} In Fig.~(\ref{oscliberaexp}) we show both the
theoretical prediction (solid line) and the experimental
observation (solid circles) of the dipole oscillation of a Fermi
gas at $T=0.3 \, T_F$ in presence of a lattice with $s$=3. For
comparison, we show also the dipole motion in the absence of the
periodic potential (dashed line and open circles) which consists
in an undamped harmonic oscillation at the trap frequency $\w_z$.
We observe a dramatic change of the dipole oscillation in presence
of the lattice, which is well described by the 3-D model. The
appearance of an offset in the oscillations is due to the
significant fraction of particles moving along open orbits; for
the given parameters the Fermi energy is indeed larger than the
bandwidth 2$\delta\approx 0.4 E_F$. This fraction behaves
macroscopically as an insulator, because its center of mass does
not move under the harmonic force but stays trapped on one side.
The fraction of the gas occupying closed orbits can instead
oscillate in the harmonic potential, and has therefore a
conducting nature. A damping appears as expected because of the
dephasing between different orbits. Also, the oscillation
frequency is reduced because of the larger effective mass of the
atoms in the lattice.
\begin{figure}
\includegraphics[scale=0.57]{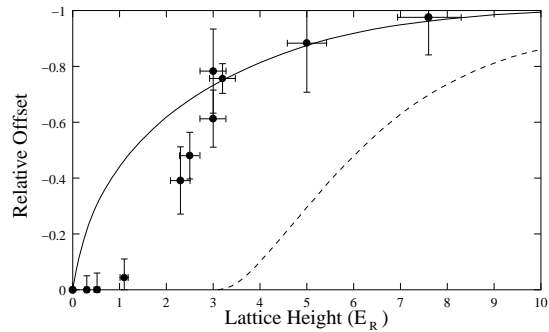}
\caption{\small{\textit{Relative oscillation} Relative offset of
the oscillations of a Fermi gas in the lattice (normalized
to the initial displacement) as a function of the lattice height.
The circles are the experimental data and the continuous
line is the theoretical prediction for 3$\times 10^4$ atoms at
$T$=100~nK. The dashed line is the prediction for a Fermi gas of
2.5$\times 10^3$ atoms at $T$=0.}}\label{trapexp_articolo}
\end{figure}
To understand how such phenomenology depends on the width of the
first energy band, we have performed a series of measurements by
keeping the atom number and temperature of the Fermi gas constant,
and varying the lattice height. In this way it was possible to
move the Fermi energy within the energy gap between the first and
second band of the lattice. In
Figs.~\ref{trapexp_articolo}-\ref{dampingexp_articolo} we plot the
measured dependencies of the offset, damping rate and oscillation
frequency, that are in good agreement with the theoretical
calculations. Fig.~\ref{trapexp_articolo} shows the crossover from
a conducting behavior in low lattices to an almost completely
insulating behavior in higher lattices. Here we have plotted the
relative oscillation offset defined as $z_{osc}/z_0$, where
$z_{osc}$ is the center of oscillation of the system in the
lattice, as a function of the lattice height. Note how the
relative offset, which represents the insulating fraction of the
Fermi gas, stays small as long as 2$\delta< E_F$, and then raises
quite rapidly towards unity. Since in the present experiment
$E_F\approx E_R$, an insulating fraction appears already with low
lattices; the theory however shows that the threshold for the
insulation moves to higher lattices in case of smaller Fermi
energies (dashed line in Fig.~\ref{trapexp_articolo}). 
The disagreement between experiment
and theory at low lattice heights, $s<3$, can arise from the
population of higher bands due to the finite temperature
and/or Landau-Zener tunnelling, which are not included in the
theoretical model. Increasing $s$, the energy gap between bands
increases, and the single band calculations become more realistic.

In Fig.~\ref{dampingexp_articolo} we show the observed features of
the conducting fraction of the gas. The damping rate of the
oscillation shown in Fig.~\ref{dampingexp_articolo}a, also
increases with the lattice height, because of an increased
dispersion of the oscillation frequencies of atoms in closed
orbits \ref{w0E}.  As shown in Fig.~\ref{dampingexp_articolo}b,
the oscillation frequency of this component of the gas is close to
that expected for a particle at the bottom of the band
$\omega_0(E_z=0)$. Actually, the theoretical analysis shows that
this value is strictly reached only asymptotically during the
oscillation, because initially closed orbits with different
frequencies contribute to the dipole motion.

As a consequence of the Pauli principle, which keeps the
energy distribution broad, the Fermi gas exhibits an insulating
behavior even at $T$=0. 
The observed phenomena only weakly depend on the
gas temperature, at least in the region 0.2-1 $T_F$ that we have
explored so far in both experiment and theory \cite{nota}, and in
general we observe an increase of both offset and damping for
increasing temperatures, as expected because of the broader energy
distribution. An analogous insulating phenomenon can be observed
in uncondensed Bose gases at temperatures larger than the
bandwidth. Indeed also in this case a
significative part of the single particle orbits will have energy
in the gap region giving rise to insulation. 
However we have observed \cite{Ott} that in the
case of bosons, or even in the case of fermions admixed with
bosons, interatomic collisions quench the open orbits and
eventually bring the system into the equilibrium position.

{\em Conclusion.} In this paper we have investigated the dynamics of a
non-interacting Fermi gas in presence of a combined periodic and
harmonic potential, comparing the experimental observation with a
semiclassical theory. We have shown that the system crosses from a
conducting to an insulating behavior by tuning the Fermi energy
relatively to the first energy band of the lattice. The
present work provides a basis for further investigation of
interacting Fermi gases in lattices, in both normal and superfluid
phases, including
the superfluid Josephson-like oscillations already observed in
a Bose-Einstein condensate confined in the same trapping potential
\cite{Cataliotti}.

This work was supported by MIUR, by EU under contract
HPRICT1999-00111, and by INFM, PRA ``Photonmatter''. H.O. was
supported by EU under contract HPMF-CT-2002-01958.



\begin{thebibliography}{100}
\bibitem{banda}
For small values of the lattice heigth we fit the band dispersion
with $\varepsilon(\Pz)=c \,\frac{\Pz^2}{2 m} \, \Big(\exp\big(a \,
(|\frac{\Pz d}{2 \hbar}|-b)\big)+1\Big)^{-1}$, where $a$, $b$, $c$
are free parameters.
\bibitem{nota1} We have checked that the observed phenomenology
does not depend on the displacement up to $z_0$=35 $\mu$m,
justifying the use of a linear approach in the theory.
\end{thebibliography}
\end{document}